\documentclass[letterpaper]{jpconf}
\usepackage{graphicx}
\usepackage{epstopdf}

\begin{document}

\title{BRAHMS Overview }

\author{Ramiro Debbe\dag\ for the BRAHMS Collaboration  }

\address{\dag\ Brookhaven National Laboratory, Upton NY, 11973 }

\begin{abstract}
A brief review of BRAHMS measurements of bulk particle production in RHIC Au+Au collisions at $\sqrt{s_{NN}}=200GeV$ is presented, together 
with some discussion of baryon number transport. Intermediate $p_{T}$ measurements in different collision systems (Au+Au, d+Au and p+p) are
also discussed in the context of jet quenching and saturation of the gluon density in Au ions at RHIC energies. This 
report also includes preliminary
results for identified particles at forward rapidities in d+Au and Au+Au collisions.

\end{abstract}




\section{Introduction}

The vast majority of particles produced in high energy hadron collisions (pp, AA) have 
transverse momenta with steeply falling distributions and consists mostly of pions. This
bulk particle production is often refered to as ``soft physics''; an up to now intractable
non-perturbative domain of QCD that remains poorly understood.
Even though these collisions can be modeled with varying degrees of accuracy, a 
solution from ``first principles'' is still being sought. The BRAHMS collaboration at RHIC 
has set out to measure the bulk particle production in as wide as possible range in rapidity 
 with good particle identification and high momentum resolution. The wide rapidity range offers
 an almost complete coverage that allows for the inclusion of conservation laws whenever a 
description of the physics behind the collisions is proposed. This contribution starts with
a brief summary of some of BRAHMS most relevant measurements of bulk particle production.

The production of jets at RHIC energies is well established and, one of the most dramatic 
discoveries in Au+Au collisions at full energy has been 
the suppression of those jets as they traverse a highly opaque medium formed by the 
collisions \cite{RHIC-AuAu-Supp, BRAHMS-Supp}.  Studies of jet production with the BRAHMS 
apparatus are done 
 through the measurement of leading particles up to intermediate values of transverse momentum
 ($\sim 4 GeV/c$).
Our ability to identify the particles we measure, together with our high  
rapidity reach, make BRAHMS intermediate $p_{T}$ 
measurements very relevant to study the longitudinal component of the new medium formed in 
Au+Au collisions. The high 
rapidity reach has also been instrumental in the study of charged particle production in d+Au
 collisions compared to incoherently added p+p collisions. The evolution of this comparison 
with rapidity and the centrality of the collision will be reviewed in this presentation.    
Finally, preliminary results from the analysis of Au+Au and d+Au collisions are presented in 
the last section. More details about BRAHMS results can be found in our ``white paper'' 
\cite{WhitePaper}.

\section{Experimental setup}
The BRAHMS setup consists of two rotatable  spectrometers, the mid-rapidity 
spectrometer (MRS) and the 
 forward spectrometer (FS), complemented with an event characterization  system used to determine the 
geometry of the collisions.
The MRS spectrometer measures the momentum of charged particles with two tracking stations (time projection chambers) and a single dipole magnet. Particle identification in this spectrometer is done with time-of-flight hodoscopes and a threshold Cherenkov detector. 
Details about the performance of the MRS spectrometer can be found in Ref. \cite{centralityDep}. 
The FS spectrometer measures the much higher momenta of charged particles produced at small angles with 
five tracking stations 
(two TPC and three drift chambers).  Particle identification in the FS spectrometer 
is done with a complement of two time-of-flight hodoscopes, one threshold Cherenkov counter and a Ring Imaging Cherenkov detector.
A detailed description of the BRAHMS experimental setup can be found in \cite{BRAHMSNIM}. 
The geometry of the collisions is extracted from the multiplicity of charged particles measured in the $\mid \eta \mid \leq 2.2$ range \cite{BRAHMS-Tiles}.
The normalization of our measurements is obtained with minimum biased triggers designed to maximize the 
coverage of the inelastic cross section. In Au+Au collisions that trigger was defined with the Zero Degree
 Calorimeters, ZDC, and for the p+p and d+Au collisions, with a set of scintillators located around the beam pipe.

\section{Rapidity densities}
\begin{figure}[!ht]
\begin{center}
\resizebox{0.8\textwidth}{!}
           {\includegraphics{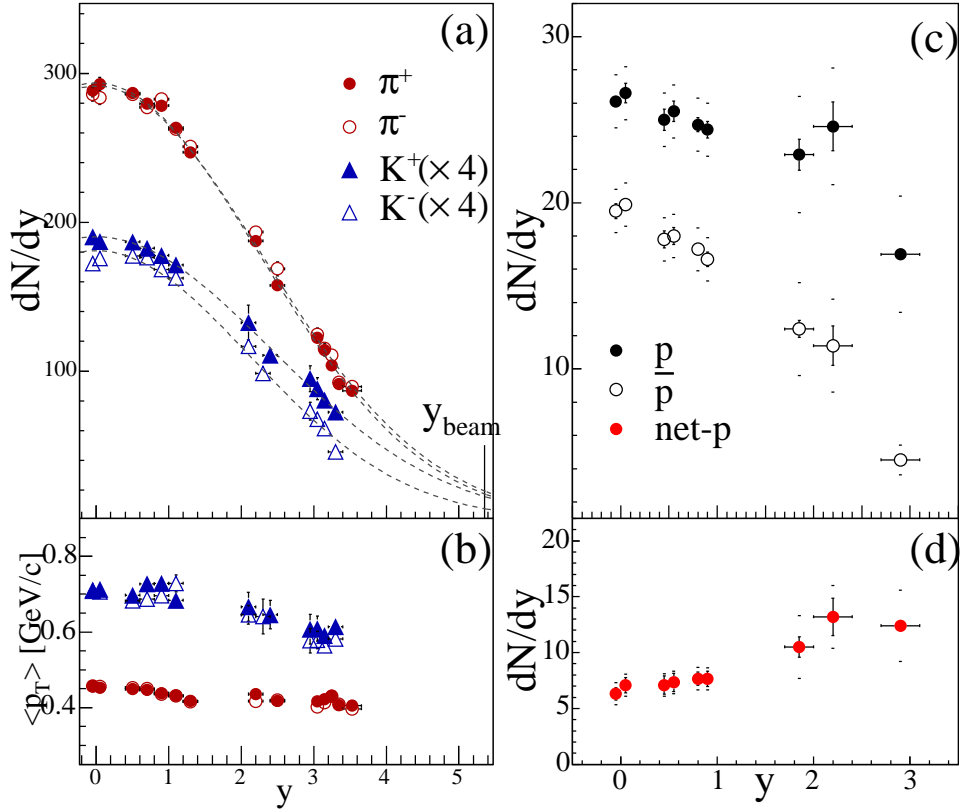}}
\end{center}
\caption{\label{fig:dndy}(a) Rapidity densities for pions and kaons. (b) Mean transverse momentum for pions, kaons
and protons. (c) Rapidity density of protons and anti-protons. (d) Net proton rapidity densities corrected for 
hyperon feed-down. Small horizontal lines are used to display the systematic errors for the measurements displayed 
in panels c and d.}
\end{figure}

The rapidity densities for the most central data sample $(0-5\%)$ in Au+Au collisions at $\sqrt{s_{NN}} = 200 GeV$ have been obtained by integration of the invariant 
differential yields $\frac{1}{2\pi} \frac{d^{2}N}{p_{T}dp_{T}dy}$ at several intervals of 
rapidity. The measurement of very low values of transverse momentum $p_{T}$ is limited by multiple 
scattering and particle decay, an interpolation is thus necessary in order to integrate the 
yields all the way down to zero $p_{T}$. Power law 
functions were used to integrate the pion distributions and single exponentials in $m_T - m_0$ 
for the kaon distributions \cite{chargedMeson}. Protons 
were fitted with single gaussians \cite{stopping}. The result of these integrals is shown in 
Fig. \ref{fig:dndy}.
The shape of the produced particle densities (pions, kaons and anti-protons) is remarkably Gaussian. The width of the negative pion distribution is 
equal to: $ \sigma_{\pi^{-}} = 2.29 \pm 0.02 $.  The widths of the rapidity densities of pions, kaons and anti-protons are very similar, 
and at the same time different from the ones associated with a single thermal source. A hydrodynamical description of the system is considered
as the best approach to explain this wide distributions, even though hydrodynamical models \cite{Hirano}, have not yet been able to reproduce the rapidity
dependence of the so called elliptic flow \cite{EllipticFlow}.  

Panel c of Fig. \ref{fig:dndy}  shows the measured rapidity density for protons and 
anti-protons and panel d shows the net-proton density as function of rapidity after 
corrections from hyperon feed down. From this measurement an average rapidity loss of
2 units of rapidity has been deduced. The fact that the net proton yield around mid-rapidity 
is considerably different from zero, suggests that there are other mechanisms, besides 
valence quark transport that can move baryon number to mid-rapidity. Baryon 
junctions are considered as good candidates because their small-x component would be a
natural way to place finite baryon number at $y \sim 0$ \cite{BaryionJunction}. More details of this measurement can be found in ref. \cite{stopping}.

With the measurements described above, there is enough information to do an energy balance of the Au+Au collisions at $\sqrt{s_{NN}} = 200 GeV$. 
With reasonable assumptions about the particles that were not measured, and allowing for some 
$10-15\%$ error in the 
extrapolation up to beam rapidity, we find that 25 TeV out of the 33 TeV of total energy is found in produced particles.
The integrated energy of the extrapolated net baryon distribution has been found to be $ 27 \pm 6\ GeV$ (see Ref. \cite{stopping}),
leaving $73 \pm 6\ GeV$, out of the 100 GeV of each incoming nucleon, available for particle 
production, well in agreement 
with the result obtained from the energy balance per particle species. 
  
\section{Intermediate $p_{T}$ studies and the nuclear modification factors}

As mentioned above, one of the most important results from RHIC is the suppression of
intermediate $p_{T}$ compared to appropriately 
scaled p+p collisions \cite{RHIC-AuAu-Supp, BRAHMS-Supp}.
Such 
reduction in yield was soon identified as energy loss in a dense and opaque medium;
partons with high fractional momentum traverse that medium while their energy is being degraded by multiple interactions
(mainly  gluon bremsstrahlung) and later hadronize into jets whose leading particles are then detected.  
 Similar measurements performed in d+Au collisions at
the same energy did not show the same strong suppression at mid-rapidity but rather
an enhancement \cite{RHICdA, BRAHMS-Supp}. This enhancement is understood as multiple elastic scatterings 
at the partonic level before the interactions that produce the jets whose leading particles are detected. The dominant source of the suppression measured in Au+Au collisions 
would then be the final state interactions with the opaque medium formed at the collision. BRAHMS extended a similar study to higher values of rapidity and found that the description
of the enhancement measured at mid-rapidity in d+Au collisions was not applicable at forward rapidities,
where in fact, we found again a suppression \cite{rdAPRL}.
\begin{figure}[!ht]
\begin{center}
\resizebox{0.8\textwidth}{!}
           {\includegraphics{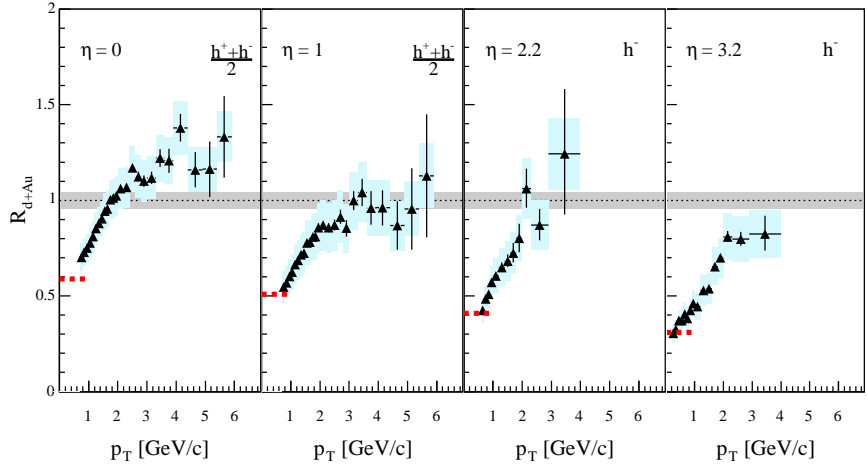}}
\end{center}
\caption{\label{fig:ratio} Nuclear modification factor for charged
  hadrons at pseudorapidities $\eta=0,1.0,2.2,3.2$. Statistical
  errors are shown with error bars. Systematic
  errors are shown with shaded boxes 
  with widths set by the bin sizes.                                          
  The 
  shaded band around
  unity indicates the estimated error on the normalization to $\langle N_{coll} \rangle$. 
  Dashed lines at $p_T<1$ GeV/c show the normalized charged particle 
  density ratio $\frac{1}{\langle
  N_{coll}\rangle}\frac{dN/d\eta(d+Au)}{dN/d\eta(pp)}$.}
\end{figure}

Figure \ref{fig:ratio} shows the nuclear modification factor defined as
$R_{dAu}=\frac{1}{N_{coll}}\frac{\frac{dN^{dAu}}{dp_{T}d\eta}}{\frac{dN^{pp}}{dp_{T}d\eta}}$, 
where $N_{coll}$ is the number of binary collisions estimated to be equal to $7.2 \pm 0.6$ for
 minimum biased d+Au 
collisions.  Each panel shows the ratio calculated at
a different $\eta$ value. 
At mid-rapidity ($\eta = 0$), the nuclear modification factor exceeds 1 for
transverse momenta greater than 2 GeV/c in a similar, although less pronounced way 
as Cronin's p+A measurements
performed  at lower energies \cite{CroninEXP}. 
 
A shift of one unit of rapidity is enough to make the Cronin type enhancement disappear, and as 
the measurements are done at higher rapidities, the ratio becomes consistently smaller than 1 
indicating a suppression in d+Au collisions compared to scaled p+p systems at the same energy. 

In all four panels, the statistical errors, shown as error bars (vertical lines), are dominant specially in our most
forward measurements. Systematic errors are shown as shaded rectangles. An additional systematic error is introduced 
in the calculation of the number of collisions $N_{coll}$ that scales the d+Au yields to a nucleon-nucleon 
system.  That error is
shown as a $15\%$ band at $R_{dAu}=1$.  

\begin{figure}[!ht]
\begin{center}
\resizebox{0.8\textwidth}{!}
           {\includegraphics{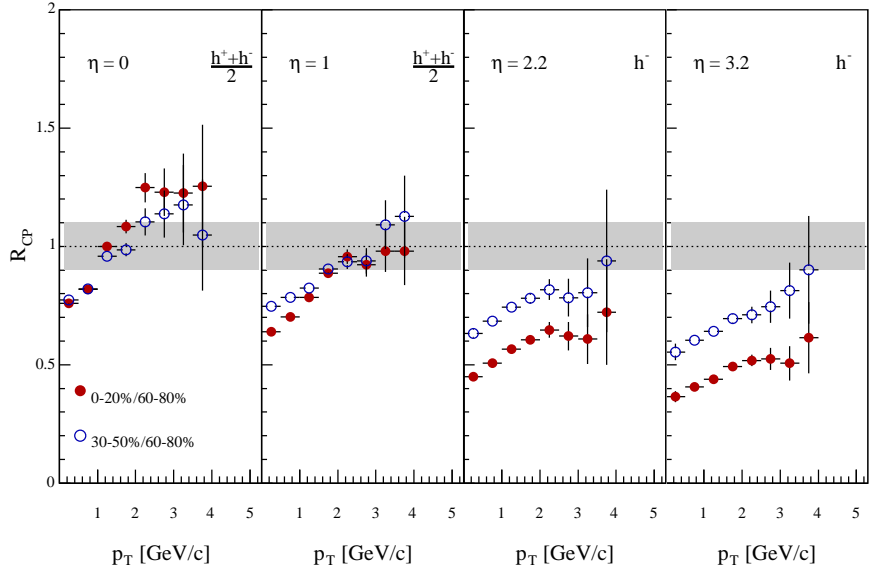}}
\end{center}
\caption{\label{fig:centrality} Central (full points) and
    semi-central (open points) $R_{cp}$ ratios (see text for details)
    at pseudorapidities $\eta=0,1.0,2.2,3.2$. Systematic errors ($\sim5\%$) 
    are smaller than the symbols. The ratios at the highest pseudorapidities ($\eta=2.2$ and 3.2) are calculated for 
negative hadrons. The uncertainty on the normalization of the ratios is displayed as a shaded band 
around unity. Its value has been set equal to the error in the calculation of $N_{coll}$ in the most peripheral collisions ($12\%$). }
\end{figure}

The four panels of Fig. \ref{fig:centrality} show the central $R^{central}_{CP}$ (filled symbols) and semi-central $R^{s
emi-central}_{CP}$
(open symbols) ratios for the four $\eta$ settings. The evolution as function of rapidity 
seen in this figure is more obvious for samples of central events.  Starting on the left panel corresponding to $\eta=0$, the 
central events yields are 
systematically higher than those of the semi-central events, but at the highest pseudo-rapidity  $\eta=3.2$, the yields of central events are 
$\sim 60\%$ lower than the semi-central events for all values of transverse momenta. 
These results have generated much interest in the community because they appear as another indication of the onset of the so called Color Glass Condensate \cite{McLerranVenu} at RHIC energies and high atomic numbers (A=179 for the gold ions) \cite{KKTandOthers}.

\begin{figure}[!ht]
\begin{center}
\resizebox{0.8\textwidth}{!}
           {\includegraphics{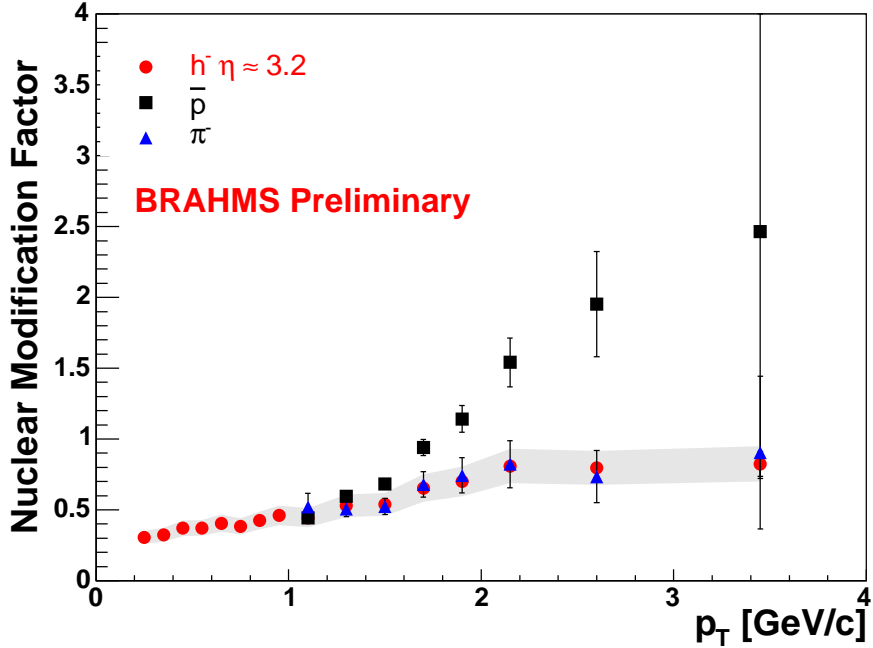}}
\end{center}
\caption{\label{fig:identifiedRdA}The nuclear modification factor $R_{dAu}$ calculated for
anti-protons (filled squares) and negative pions (filled triangles) at $\eta = 3.2$. The same 
ratio calcutated for negative particles at the same pseudo-rapidity \cite{rdAPRL} is shown with filled circles, 
and the systematic error for that
measurement is shown as grey band. }
\end{figure}

The analysis of the BRAHMS data collected from d+Au and p+p collisions is currently in progress, in particular, Fig. \ref{fig:identifiedRdA} presents the minimum bias nuclear modification $R_{dAu}$ for anti-protons and negative pions at $\eta = 3.2$. The extraction of these ratios involved the following assumptions: at each $p_{T}$ bin, one can extract
 the nuclear modification factor for identified particles, by multiplying the numerator and the denominator of this factor 
by the fractions of raw counts of identified particles to raw counts of negative particles for d+Au and p+p systems respectively: 

\[ R^{\bar{p}}_{dAu} = 
R^{h^-}_{dAu} \frac{(\frac{\bar{p}}{h^-})^{dAu}}{(\frac{\bar{p}}{h^-})^{pp} } = 
\frac{1}{N_{coll}}\frac{\left.\frac{dn^{dAu}}{dp_{T}d\eta}\right)^{h^{-}}}{\left.\frac{dn^{pp}}{dp_{T}d\eta}\right)^{h^{-}}} \frac{\frac{\left.\frac{dn^{dAu}}{dp_{T}d\eta}\right)^{\bar{p}}}{\left.\frac{dn^{dAu}}{dp_{T}d\eta}\right)^{h^{-}}}}
{\frac{\left.\frac{dn^{pp}}{dp_{T}d\eta}\right)^{\bar{p}}}{\left.\frac{dn^{pp}}{dp_{T}d\eta}\right)^{h^{-}}}} = 
\frac{1}{N_{coll}}\frac{\left.\frac{dn^{dAu}}{dp_{T}d\eta}\right)^{\bar{p}}}{\left.\frac{dn^{pp}}{dp_{T}d\eta}\right)^{\bar{p}}} \]

The ratio of raw counts is then equated with the ratio of differential yields in $\eta $ and 
$p_{T}$, assuming that all corrections do cancel out. The ratio of raw counts was
obtained with information from the Ring Imaging Cherenkov detector whose efficiency is high 
($\sim 95\%$) but has not been included in this analysis.
The errors shown for the nuclear modification factors of anti-protons and pions include the contributions from
correlations between the parameters
of the fits made to the ratios of raw counts. No attempt was made to estimate the contributions from anti-lambdas feed down 
to the 
anti-proton result. The remarkable difference between 
baryons and mesons has also been seen at RHIC energies around mid-rapidity, and has been related to parton recombination
\cite{RudyHwa}. The same explanation could be offered for these high rapidity results, but it is difficult to imagine the existence of the thermal bath of partons at high rapidity in d+Au collisions. 

\begin{figure}[!ht]
\begin{center}
\resizebox{0.8\textwidth}{!}
           {\includegraphics{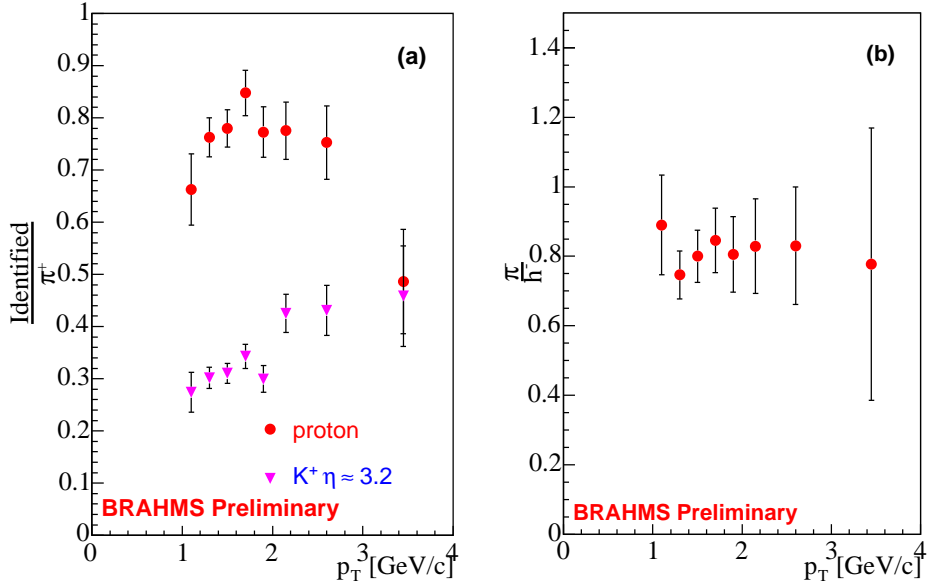}}
\end{center}
\caption{\label{fig:identifiedRatiosdA}Some results with pid at $\eta = 3.2$: (a) Particle composition of positive
charged hadrons at $\eta=3.2$ (b) The fraction of negative hadrons identified as pions at the same pseudo-rapidity. }
\end{figure}

At the time the invariant yields at $\eta = 3.2$ where shown for the first time  \cite{QM2004}, the clear 
difference between positive and negative charged particles, prompted some to associate the
suppression seen in d+Au collisions at forward angles with the so called ``beam fragmentation''
 \cite{RIKEN}. More recent work based on NLO pQCD has indicated the difficulty in reconciling the data with 
the behavior
of standard fragmentation functions \cite{Strickman}. Panel a of Fig. \ref{fig:identifiedRatiosdA} shows that, 
there are 
as many protons as positive pions ($\sim 80\%$ at $p_{T} \sim 2 GeV/c$) in the positive particle distribution at 
$\eta = 3.2$. The abundance of baryons at this high energy and rapidity doesn't support the idea of baryon 
suppression in the fragmentation region \cite{Dumitru, Strikman2} where, because of their high energy, the quarks 
of the beam would fragment independently mostly into mesons.  Panel b of the same figure shows that for all values of $p_{T}$, the fraction of 
negative pions is high and equal to $80\%$.

\begin{figure}[!ht]
\begin{center}
\resizebox{0.8\textwidth}{!}
           {\includegraphics{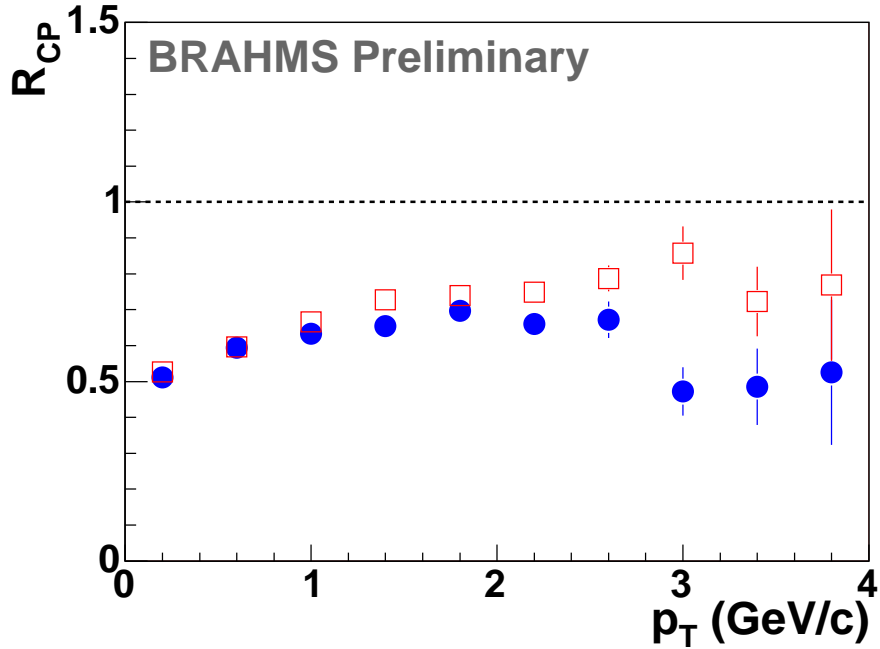}}
\end{center}
\caption{\label{fig:RcpAtHighEta} Rcp of charged particles from Au+Au at $\eta = 3.2$ Open squares: positives, filled circles: negatives.}
\end{figure}

The suppression of particle production at mid-rapidity in Au+Au central collisions compared 
to incoherently 
added p+p interactions,
or properly scaled disributions obtained from peripheral Au+Au collisions, is widely accepted as a necessary 
condition for momentum degradation by gluon radiation, as partons  traverse 
an opaque medium before they hadronize. That momentum degradation is expected to be a function of the length of the 
parton's path inside the medium as well as the medium density \cite{WangELoss}. In the above mentioned work, such 
partonic energy loss is encoded in a medium 
modified fragmentation function that has the standard (in vacuum) behavior, but only after the energy has been 
degraded by several interactions in the medium. These interactions are distributed in number according to a 
Poisson distribution with a 
mean value equal to $L/\sigma \rho$ where L is the parton's path length in the medium, $\sigma $ is the 
interaction cross-section and $\rho$ the medium density. The final fragmentation also includes the gluons 
radiated at each interaction. The pion rapidity density, which must be directly related to the medium density,
changes by almost a factor of three  between mid-rapidity and y = 3 (see Fig. \ref{fig:dndy}). 
 One would then expect less energy loss at y=3, by a comparable factor, but as can be
seen in Fig. \ref{fig:RcpAtHighEta}, there is no noticeable difference between the $p_{T}$ 
suppressions at $\eta = 3.2$ and previous measurements at mid-rapidity and $\eta=2.2$ \cite{BRAHMS-Supp}. This result
may indicate the interplay between energy loss effects that get weaker at higher rapidities, and initial state related 
suppression that becomes stronger as the number of gluons in one of the Au ions is further reduced by gluon fusion. 
On the other hand, if the $p_{T}$
distributions at forward rapidities are stepper than the ones measured at mid-rapidity, smaller energy losses would be
magnified producing similar results as the measurements. Further analysis on this subject is currently in progress.

In summary, BRAHMS has studied the properties of bulk particle production as well as baryon number transport
in Au+Au collisions at $\sqrt{s_{NN}} = 200 GeV/c$. We have also compared charged particle production in Au+Au
and d+Au collisions to similar production in p+p collisions at the same energy. Such comparisons, show an strong
suppression at intermediate transverse momentum that is associated with the formation of a dense and opaque medium.
In d+Au collisions at the same energy, such suppression does not appear at mid-rapidity, but is present at forward rapidities
and is even more pronounced in central collisions.

\section{Acknowledgments}

This work was supported by 
the Office of Nuclear Physics of the U.S. Department of Energy, 
the Danish Natural Science Research Council, 
the Research Council of Norway, 
the Polish State Committee for Scientific Research (KBN) 
and the Romanian Ministry of Research.

\end{document}